\documentclass[twocolumn,showpacs,aps,prl]{revtex4-1}
\usepackage{verbatim} %for comment
\usepackage{graphicx} % Include figure files
\usepackage{dcolumn}  % Align table columns on decimal point
\usepackage{bm}       % bold math
\usepackage{array}
\usepackage{booktabs}
\usepackage{amsmath}
\usepackage{color}
\usepackage{times}
\usepackage{epstopdf}

\usepackage[T1]{fontenc}

\begin{document}
\title {\boldmath
First Observation of a Three-Resonance Structure in $e^+e^-\rightarrow${Nonopen} Charm Hadrons
}
\author{
\begin{small}
\begin{center}
M.~Ablikim$^{1}$, M.~N.~Achasov$^{5,b}$, P.~Adlarson$^{75}$, X.~C.~Ai$^{81}$, R.~Aliberti$^{36}$, A.~Amoroso$^{74A,74C}$, M.~R.~An$^{40}$, Q.~An$^{71,58}$, 
Y.~Bai$^{57}$, O.~Bakina$^{37}$
% ...
\\
\vspace{0.2cm}
(BESIII Collaboration)\\
\vspace{0.2cm} {\it
$^{1}$ Institute of High Energy Physics, Beijing 100049, People's Republic of China\\
% ...
$^{5}$ Budker Institute of Nuclear Physics SB RAS (BINP), Novosibirsk 630090, Russia\\
% ...
$^{36}$ Johannes Gutenberg University of Mainz, Johann-Joachim-Becher-Weg 45, D-55099 Mainz, Germany\\
$^{37}$ Joint Institute for Nuclear Research, 141980 Dubna, Moscow region, Russia\\
% ...
$^{40}$ Liaoning Normal University, Dalian 116029, People's Republic of China\\
% ...
$^{57}$ Southeast University, Nanjing 211100, People's Republic of China\\
$^{58}$ State Key Laboratory of Particle Detection and Electronics, Beijing 100049, Hefei 230026, People's Republic of China\\
% ...
$^{71}$ University of Science and Technology of China, Hefei 230026, People's Republic of China\\
% ...
$^{74}$ University of Turin and INFN, (a)University of Turin, I-10125, Turin, Italy; (b)University of Eastern Piedmont, I-15121, Alessandria, Italy; (c)INFN, I-10125, Turin, Italy\\
$^{75}$ Uppsala University, Box 516, SE-75120 Uppsala, Sweden\\
% ...
$^{81}$ Zhengzhou University, Zhengzhou 450001, People's Republic of China\\
\vspace{0.2cm}
% ...
$^{b}$ Also at the Novosibirsk State University, Novosibirsk, 630090, Russia\\
% ...
}
\end{center}
  \vspace{0.4cm}
  \end{small}
}
\noaffiliation
\vspace{0.0cm}

\begin{abstract}
We report the measurement of the inclusive cross sections for $e^+e^-$$\rightarrow${nOCH} (where nOCH denotes non-open charm hadrons)
with improved precision at center-of-mass (c.m.) energies from 3.645 to 3.871 GeV. We observe three resonances: $\mathcal R(3760)$, 
$\mathcal R(3780)$, and $\mathcal R(3810)$ with significances of $8.1\sigma$, $13.7\sigma$, and $8.8\sigma$, respectively. 
The $\mathcal R(3810)$ state is observed for the first time, while the $\mathcal R(3760)$ and $\mathcal R(3780)$ states are observed
for the first time in the nOCH cross sections. Two sets of resonance parameters describe the energy-dependent line shape of the cross sections well.
In set I [set II], the $\mathcal R(3810)$ state has mass $(3805.7 \pm 1.1 \pm 2.7)$ [$(3805.7 \pm 1.1 \pm 2.7)$] MeV/$c^2$, total width 
$(11.6 \pm 2.9 \pm 1.9)$ [$(11.5 \pm 2.8 \pm 1.9)$] MeV, and an electronic width multiplied by the nOCH decay branching fraction
of $(10.9\pm 3.8\pm 2.5)$ [$(11.0\pm 3.4\pm 2.5)$] eV. In addition, we measure the branching fractions 
${\mathcal B}[{\mathcal R}(3760)$$\rightarrow${nOCH}$]=(25.2 \pm 16.1 \pm 30.4)\% [(6.4 \pm 4.8 \pm 7.7)\%]$ and 
${\mathcal B}[\mathcal R(3780)$$\rightarrow${nOCH}$]=(12.3 \pm 6.6 \pm 8.3)\% [(10.4 \pm 4.8 \pm 7.0)\%]$ for the first time. 
The $\mathcal R(3760)$ state can be interpreted as an open-charm (OC) molecular state, but containing a simple four-quark state component. 
The $\mathcal R(3810)$ state can be interpreted as a hadrocharmonium state.
\end{abstract}

\maketitle 

Until two decades ago, it was widely believed that hadron resonances with masses
higher than the open-charm (OC) pair thresholds decay entirely to OC final states 
via the strong interaction. 
However, in July, 2003, BES reported the observation of seven 
events of  hadron resonance(s) in this mass regime decaying to 
nOCH~\cite{hep_ex_0307028v1, 
Psi3770AndItsRelatedPhysicsAtBESII_2019,
PhysRevD102_112009_Y2020,
PRL127_082002_Y2021}.  
This discovery overturned the understanding of resonance decays 
and opened up a new era in hadron spectroscopy.
In this Letter, we denote these resonances as 
$X_{\rm aboveOC}$~\cite{PhysRevD102_112009_Y2020},
which encompasses both supposed pure quark-antiquark $c\bar{c}$ states, 
i.e. $\psi(3770)$, $\psi(4040)$, $\psi(4160)$, and $\psi(4415)$, 
and nonpure quark-antiquark $c\bar{c}$ states 
(hereafter referred to as `non-$c\bar{c}$'), such as four-quark states, 
OC-pair molecular states, 
hadrocharmonium
states, and hybrid charmonium 
states~\cite{F_E_Close_PLB578_119_y2004, 
M_B_Voloshin_JETP_23_333_y1976, 
HQ_EPJC_72_1534_y2011, 
Voloshin_PLB66_344_y2008}.
Quantum chromodynamics (QCD) expects that the non-$c\bar{c}$ states
exist in nature, and thus a discovery of these systems would be 
an important validation of the QCD predictions. 
The first nOCH final-state decay of a $X_{\rm aboveOC}$ resonance 
to be observed after the 2002 discoveries was  
$J/\psi{\pi^+}{\pi^-}$~\cite{Psi3770AndItsRelatedPhysicsAtBESII_2019,hep_ex_0307028v1,PLB605_Y2005_63},
seen by BES-II.
This final state can originate from a $c\bar{c}$ state, a non-$c\bar{c}$ state, 
or a combination of both~\cite{PRL127_082002_Y2021}.
This discovery stimulated strong interest in using $J/\psi\pi^+\pi^-$ or similar final states 
as golden channels to probe other nOC decays from the $X_{\rm aboveOC}$, 
and led to the discovery of several exotic  
states~\cite{X3872_PhysRevLett91_262001_Y2003, X4260_PRL95_142001_Y2005,
X4260_PRD74_091194_In2006,
X4260_PRL96_162003_In2006,
PRL110_252001_In2013,
PRL112_092001_In2013,
PRL111_242001_In2013,
PRL99_142002_In2007,
PRL98_212001_In2005,
PRL118_092001_2017,
PRL118_092002_2017}.

Subsequent studies of the $\psi(3770)$ resonance showed that it has a branching fraction of about $15\%$  
into nOCH final states~\cite{PLB641_145_2006,PRL97_121801_2006,PRD76_122002_2007,PLB659_74_2008}.
This fact indicates the contribution of some undiscovered states~\cite{RongG_CPC_34_778_Y2010} 
with masses around 3.773 GeV/$c^2$, which predominantly decay into nOCH. 
In 2008, the BES-II experiment observed for the first time a double-peaked structure 
named the $\mathcal Rs(3770)$ in $e^+e^-$$\rightarrow${\rm hadrons} at c.m. energies around 
{3.76~GeV}~\cite{bes2_prl_2structures}, which is composed of two states 
labeled the $\mathcal R(3760)$ and $\mathcal R(3780)$. 
The existence of the $\mathcal R(3760)$ state 
was confirmed in $e^+e^- \rightarrow J/\psi{X}$ 
by BESIII~\cite{PRL127_082002_Y2021}.
It is seen that the study of the inclusive nOCH decays of $X_{\rm aboveOC}$ 
both helps the understanding of known states, and is a sensitive probe for undiscovered resonances, 
particularly non-$c\bar{c}$ states.

In this Letter, we report a measurement of the cross sections 
for $e^+e^-\rightarrow${nOCH} at c.m. energies from 3.645 to 3.871~GeV, 
studies of $\mathcal R(3760)$ and $\mathcal R(3780)$ production and decays,   
and a search for an additional resonant state in this energy region. The data samples used in this analysis were collected 
at 42 c.m.\ energies in 2010 and correspond to a total 
integrated luminosity of 75.5 pb$^{-1}$.

The BESIII detector~\cite{bes3} response is studied using Monte Carlo (MC) samples. 
The simulations are performed using a {\sc Geant4}-based~\cite{geant4} software package.
Simulated samples for $q\bar q$ vector states 
(i.e. $u\bar u$, $d\bar d$, $s\bar s$, and $c \bar c$) and their decays to
hadrons are generated using the MC event generators {\sc kkmc}~\cite{kkmc}, 
{\sc evtgen}~\cite{BesEvtGen}, and {\sc lundcharm}~\cite{LundCharm_CJC_et_al_2000}.
Background sources are estimated with MC samples generated using {\sc kkmc},
and the MC event generators 
{\sc babayaga}~\cite{babayaga} and {\sc twogam}~\cite{twogam}.

    We select the inclusive nOCH events from the events with charged
particles or charged and neutral particles.
In order to reject background contributions from $e^+e^-\rightarrow ({\gamma})e^+e^-$ and
$e^+e^-\rightarrow ({\gamma})\mu^+\mu^-$, we require 
the events to have more than two charged tracks ($N_{\rm CTrk}$), 
and impose the following selection criteria for each track:
(i) the distance (${R_{xy}}$)
of the point of the closest approach to the beam pipe 
must satisfy the condition ${R_{xy}}\le1.0$ cm;
(ii)  the polar angle $\theta$ must satisfy
the condition $|\!\cos\theta|$$<0.93$;
(iii) the momentum $p$ must be less than 
      $E_{\rm b}$+$0.02E_{\rm b}$$\sqrt{1+E^{2}_{\rm b}}$,
      where $E_{\rm b}$ is the beam energy;
(iv)  the time-of-flight $t_{\rm TOF}$ must satisfy 
      2.0$<t_{\rm TOF}$$<$20.0 ns and $|t_{\rm TOF}$-$t_{\rm p}|$$<2.0$ ns,
      where $t_{\rm p}$ is the expected time-of-flight of protons;
(v)  the energy $E_{\rm EMC}$ deposited in the 
     electromagnetic calorimeter (EMC) must
     be less than $1$ GeV; 
(vi) the penetration depth in the muon-chamber system
     must be less than $30$ cm.

    For the selection of photons, we require the deposited energy of a neutral cluster 
in the EMC to be greater than 25~MeV in the barrel and 50~MeV in the end caps. 
We do not apply any requirements on the number of photons in the event.
To suppress electronic noise and showers unrelated to the event, 
we impose the condition that the difference between the EMC time and the event 
start time be within [0,700] ns. To reduce the beam-associated events (beam interactions 
with gas or material), we demand that at least one charged track or photon must point into 
each hemisphere of $\cos\theta<0$ and $\cos\theta>0$. In addition, for each event, 
we require the total energy ($E_{\rm EMC}^{\rm tot}$) deposited in the EMC
by the charged and neutral particles to be greater than $0.28E_{\rm b}$.

\begin{figure}[]
\centering
\includegraphics[width=0.48\textwidth,height=0.28\textwidth]{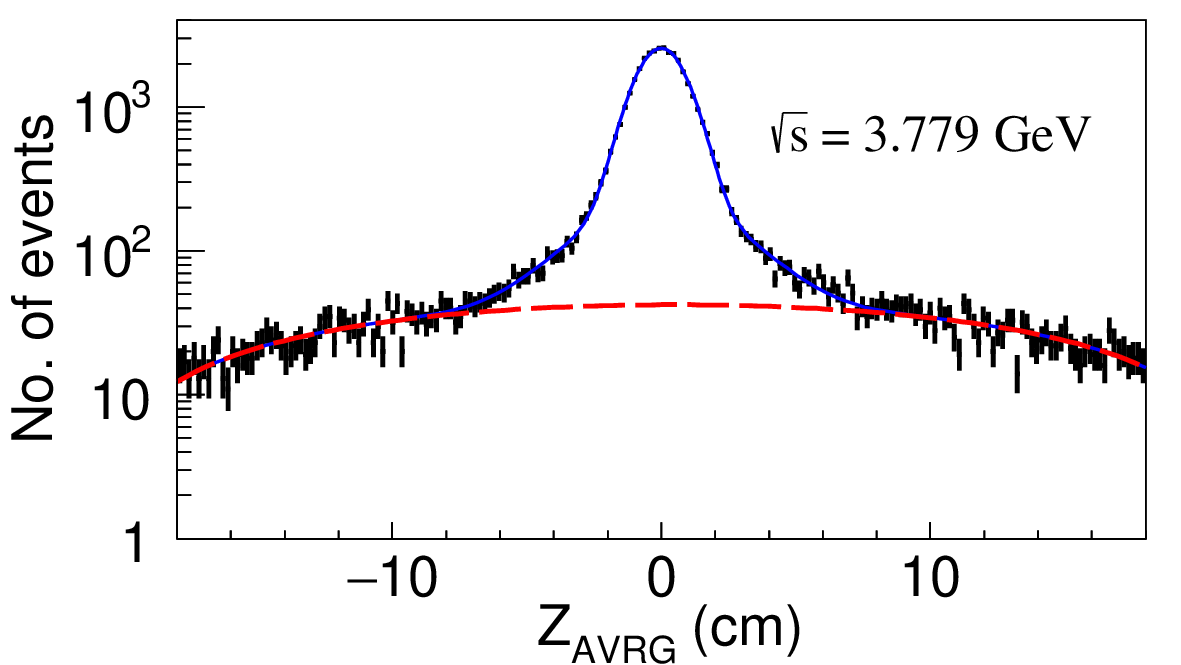}
\caption{ $Z_{\rm AVRG}$ distribution of the vertices of the selected events 
from the data sample collected at $\sqrt{s}=3.779$ GeV, where the dots with error bar 
represent data, the solid line in blue is the best fit, and the dashed line in red is 
the fitted background shape.}
 \label{fig:AvrgZ_hadEvents}
\end{figure}

Some beam-associated background sources still survive
this selection. These background sources are produced at random $z$ positions~\cite{PLB641_145_2006}, 
while genuine nOCH events are produced around $z=0$, where $z$ is the distance 
to the interaction point along the BESIII axis. To distinguish the nOCH events from the 
background sources, we calculate the averaged $z$ ($Z_{\rm AVRG}$) of the charged tracks in each event.
Figure~\ref{fig:AvrgZ_hadEvents} shows the distribution of the averaged $Z_{\rm AVRG}$ 
of the accepted events from the data sample collected at $\sqrt{s}=3.779$ GeV. Using a double-Gaussian 
function to describe the signal shape plus a second-order Chebychev function to parameterize
the background shape, we fit the $Z_{\rm AVRG}$ distribution of event vertices to extract 
the number of nOCH candidates, $N^{\rm fit}_{\rm had}$ at each energy point.

The background to the $N^{{\rm fit}}_{\rm had}$ distribution comes from various sources, e.g. 
(i)   $e^+e^-\rightarrow(\gamma)e^+e^-$,
(ii)  $e^+e^-\rightarrow(\gamma)\mu^+\mu^-$,
(iii) $e^+e^-\rightarrow\gamma\gamma$,
(iv) $e^+e^-\rightarrow(\gamma)\tau^+\tau^-$, 
(v)  $e^+e^-\rightarrow(\gamma)e^+e^-\ell^+\ell^-$ ($\ell=e$, $\mu$ or $\tau$),
(vi)   $e^+e^-\rightarrow(\gamma)e^+e^-X_{\rm had}$ (where $X_{\rm had}$ denotes hadrons), and 
(vii)  $e^+e^-\rightarrow(\gamma)$$D\bar{D}$.
The total amount of background $N_{\rm b}$  at each c.m. energy $\sqrt{s}$ is determined by 
$N_{\rm b} = \sum^{i=7}_{i=1}N_{{\rm b},i}$ with $N_{{\rm b},i}=\mathcal L\times\sigma_{{\rm b},i}\times\eta_i$,
where $\mathcal L$ is the integrated luminosity of the data sample,
$\sigma_{{\rm b},i}$ is the cross section for the $i${\rm th} background 
source and $\eta_{i}$ is the probability of misidentifying a candidate from the 
$i${\rm th} background source as a nOCH event, which is determined 
by analyzing the large background MC samples. The cross sections
$\sigma_{\rm b}$ for  sources (i), (ii) and (iii) are taken from {\sc babayaga}. For  
 source (iv), $\sigma_{\rm b}$ is calculated using the  formulae 
given in Refs.~\cite{calc_ditau1,calc_ditau2}. The cross sections 
for sources (v) and (vi) are taken from {\sc twogam},
while that for source (vii) is taken from the 
observed cross sections $\sigma^{\rm o}_{D\bar{D}}(s)$ 
for $e^+e^-$$\rightarrow$$D\bar{D}$ determined using the same data samples.
For example, at $\sqrt{s}=3.7731$ GeV, $N^{\rm fit}_{\rm had}=(35235\pm{225})$,
$N_{{\rm b},{\ell^+\ell^-, \gamma\gamma}}=(1206\pm{7})$,
$N_{ {\rm b}, {e^+e^-{\ell^+\ell^-},{e^+e^-{\rm X}_{\rm had}}} }=(341\pm{6})$, and 
$N_{ {\rm b},{D\bar{D}} }=(9458\pm 186)$,
where the uncertainties on the first two backgrounds arise from the
statistical uncertainties on $\mathcal L$, $\sigma_{{\rm b},{i}}$, and $\eta_{i}$,
and that on the third is due to the statistical uncertainty 
on $\sigma^{\rm o}_{D\bar{D}}(s)$.
To provide the most conservative signal significance for the resonance searches that are discussed below, 
we directly subtract $N_{{b}}$ 
from $N^{{\rm fit}}_{\rm had}$ and allow the statistical uncertainty 
of  $N_{ {\rm b},{D\bar{D}}}$  to fully contribute to the
uncertainty in the difference between $N^{{\rm fit}}_{\rm had}$ and $N_{{b}}$.
This procedure yields $N^{{\rm o}}_{\rm nOCH}=(24230\pm {292})$, where the uncertainty is statistical 
and  includes the contribution from the background estimates.

\begin{figure}[]
\centering
\includegraphics[width=0.48\textwidth,height=0.28\textwidth]{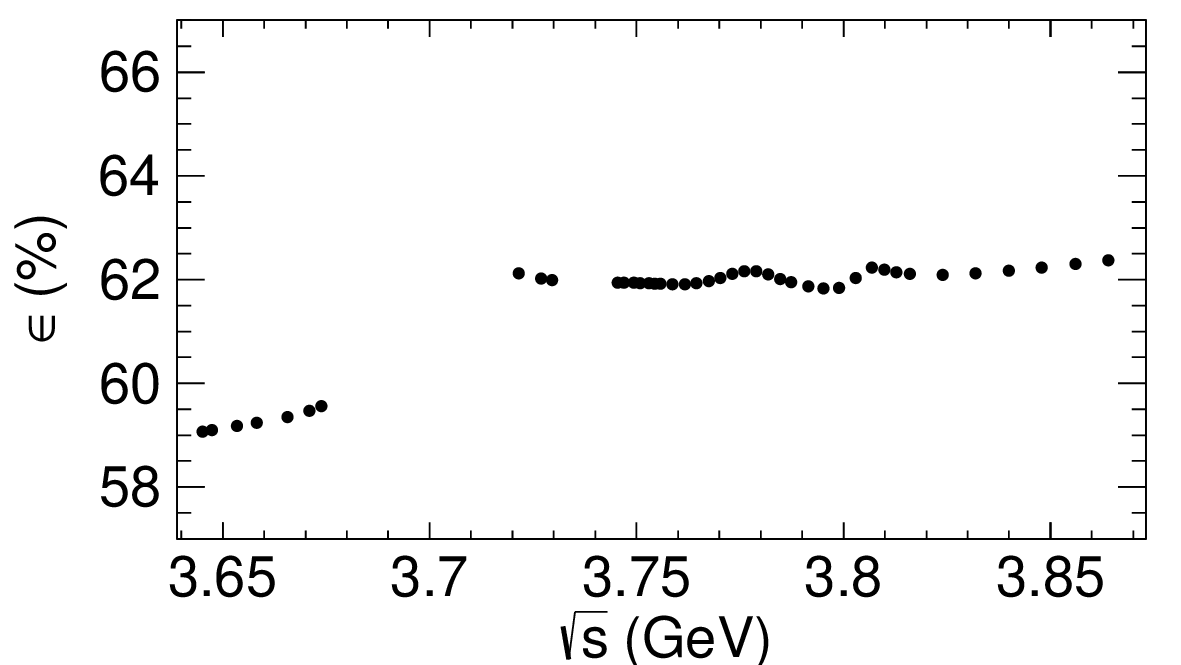}
\includegraphics[width=0.48\textwidth,height=0.28\textwidth]{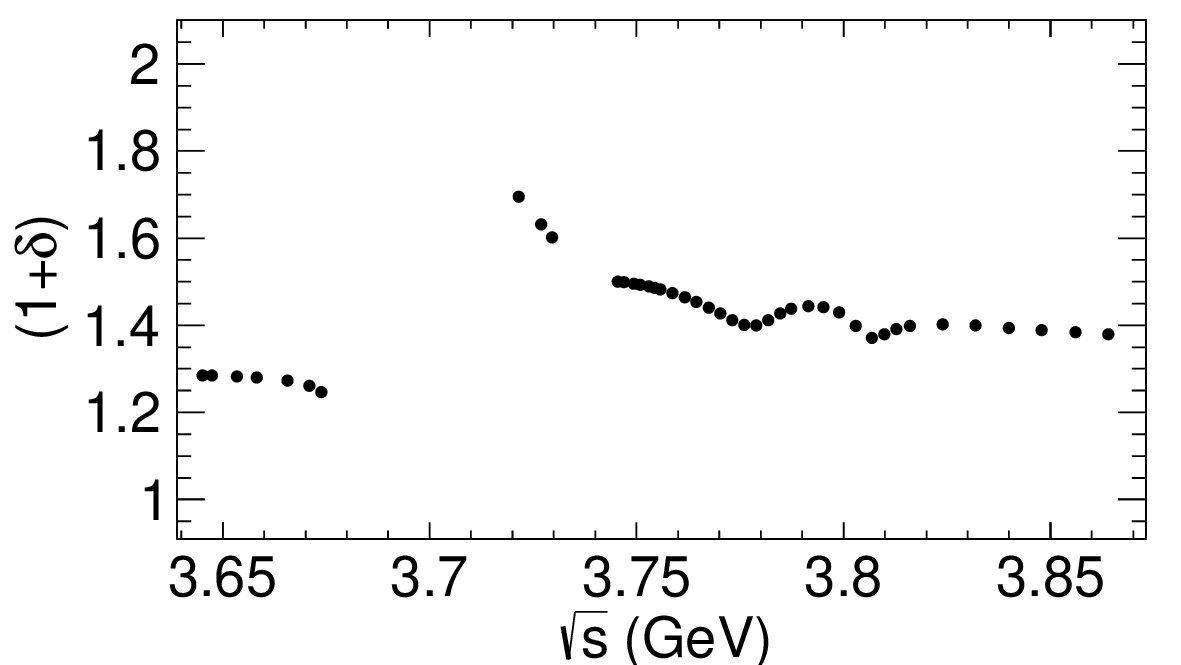}
\caption{
The efficiency $\epsilon$ (top) versus the c.m energy; the ISR correction factor
$[1+\delta(s)]$ (bottom) versus the c.m energy.}
\label{fig:eff_IstCrrF_nOCH}
\end{figure}

We determine the detection efficiency using MC simulated events 
for the four components of the process $e^+e^-$$\rightarrow\,${nOCH}:
(i) $e^+e^-$$\rightarrow\,$light-hadron (LH)  continuum processes
including lower-mass resonances (LMRs) with masses below 2 GeV (CPLMRs)
(ii) $J/\psi $$\rightarrow\,${hadrons}, 
(iii) $\psi(3686) $$\rightarrow\,${hadrons}, and
(iv) ${\mathcal Rs(3770)}$$\rightarrow\,${nOCH}~\cite{bes2_prl_2structures}.
 We generate  simulated samples for these events with the {\sc kkmc} package.  
These events include initial-state radiation (ISR)
and final-state radiation (FSR) processes.
For the subsequent decays of the $J/\psi$, $\psi(3686)$ and ${\mathcal Rs(3770)}$, 
we use {\sc evtgen} to generate the 
known final states, given in Ref.~\cite{PDG2022},
and use {\sc lundcharm} to generate the  
remaining unknown final states.
The resulting selection efficiency is determined using $\epsilon=\sum^{4}_{1}w_{i} \epsilon_{i}$, 
where $w_{i}$ is the number of nOCH events simulated for the 
$i${\rm th} component over the total number of MC simulated nOCH events for all 
four components, and
$\epsilon_{i}$ is the corresponding efficiency for selection of the nOCH events 
from the $i${\rm th} component. Figure~\ref{fig:eff_IstCrrF_nOCH} (top) shows the 
efficiencies determined at the 42 energy points.

At $\sqrt{s}=3.7731$ GeV, the efficiency is $\epsilon=62.11\%$.
The integrated luminosity corresponding to the data collected at this energy is 
$\mathcal L=(1831.63\pm 4.49)$~nb$^{-1}$.
Dividing the number $N^{{\rm o}}_{\rm nOCH}=(24230\pm {292})$
by both the integrated luminosity and the efficiency yields the observed nOCH cross
section $\sigma^{\rm o}_{\rm nOCH}=(21.299\pm 0.262)$~nb, where the uncertainty
arises from the statistical uncertainty on the $N^{\rm o}_{\rm OCH}$, 
the size of the MC samples, and the statistical uncertainty of the luminosity measurement.
Similarly, we measure $\sigma^{\rm o}_{\rm nOCH}(s)$ at
the remaining 41 energy points.

Table~\ref{table:syserr} summarizes the systematic uncertainties assigned to 
the $\sigma^{\rm o}_{\rm nOCH}(s)$ measurements.
To determine the systematic uncertainties 
due to the choice of the selection criteria, we vary each criterion from its baseline value  
to an alternative setting, as given in Table~\ref{table:syserr}, and assign  the resulting 
change in $\sigma^{\rm o}_{\rm nOCH}(s)$ as the systematic uncertainty. 
Removing the requirement that tracks must point into different hemispheres 
in $z$
changes $\sigma^{\rm o}_{\rm nOCH}(s)$ by $0.80\%$, which is taken as 
the corresponding uncertainty associated with this possible source of bias. 
To estimate the uncertainty associated with the fit to the $Z_{\rm AVRG}$ distribution, 
we reperform the fit changing the background shape from a second-order to third-order, 
and then from a third-order to fourth-order Chebychev function.
These result in $0.83\%$ change in the cross sections, which is assigned 
as the corresponding systematic uncertainty. 
The uncertainty on the number of background sources
is dominated by the estimation of the two-photon contribution, 
which induces a $0.75\%$ uncertainty of the $\sigma^{\rm o}_{\rm nOCH}(s)$. 

The choice of the MC generator for $e^+e^- \rightarrow {\rm nOCH}$ 
impacts the selection efficiency. We recalculate the efficiency with different MC 
packages and take the $1.70\%$ variation observed as the associated uncertainty. 
There is an uncertainty on $\sigma^{\rm o}_{\rm nOCH}(s)$ of $1.00\%$ 
arising from the corresponding uncertainty on the luminosity. 
Adding these contributions in quadrature yields a total systematic uncertainty 
of $2.89\%$ on  $\sigma^{\rm o}_{\rm nOCH}(s)$.
This total does not include an energy-dependent uncertainty 
on $\sigma^{\rm o}_{\rm nOCH}(s)$, caused by the $1.43\%$
uncertainty on the $\sigma^{\rm o}_{D\bar{D}}(s)$ shape, which is accounted for 
when considering the systematic uncertainties on the fitted parameter values, as discussed below. 
\begin{table}[htbp]
\renewcommand\arraystretch{1.2}
\centering
\caption{Systematic uncertainties on $\sigma^{\rm o}_{\rm nOCH}(s)$.}
\begin{tabular}{lcr}
\hline
\hline
 Source              & Variation range       & Uncertainty (\%) \\
\hline
\textbf{${R_{xy}}$} cut   & [0.9, 1.1]                &  0.20 \\
cos$\theta$ cut               & [0.875, 0.930]        &  0.40 \\
$p<$$E_{\rm b}$+0.02$E_{\rm b}$$\sqrt{1+E_{\rm b}}$   & [nominal value, $\infty$]       & 0.01       \\
$\rm t_{TOF}$ cut and ${\rm t_{TOF}}$-${\rm t_{p}}$   & [0, $\infty$] and [2,$\infty$]  &  0.09 \\
$E_{\rm EMC}$ cut                                     & [1, $\infty$] &  0.39 \\
Penetration depth                                     & [30, $\infty$] &  0.04 \\
$E_{\rm EMC}^{\rm tot}$ cut                           & [0,0.28]                        & 0.34 \\
$N_{\rm CTrk}$ cut                                    & [2,3]                           & 1.45 \\
Different hemisphere in $z$                           &                                 &  0.80 \\
Fitting $Z_{\rm AVRG}$ distribution                   &                                 &  0.83 \\
$N_{\rm b}$                                           &                                 &  0.75 \\
MC signal model                                       &                                 &  1.70 \\
Integrated luminosity                                 &                                 &  1.00 \\   
                                                      &                                 &       \\
{Total}                                               &                                 &  {2.89} \\
\hline \hline
\end{tabular}
\label{table:syserr}
\end{table}

To investigate whether the $\mathcal R(3760)$ and $\mathcal R(3780)$ states 
decay to nOCH and if a new resonance $\mathcal R$ exists in this energy region,
we perform a least-$\chi^2$ fit to the nOCH cross sections. 
The dressed nOCH cross section is modeled by:
\begin{equation} 
\label{eq:exp_D_xs_nOCH}
\sigma^{\rm D}_{\rm nOCH}(s') = \sigma^{\rm D}_{\rm LH}(s')
      + \sigma^{\rm D}_{J/\psi}(s')
      + \sigma^{\rm D}_{{\mathcal R}_{{\mathcal S}\rm up{3680}}}(s'),
\end{equation}
where
$s'=s(1-x)$, $x$ is the radiative-photon energy fraction,
$\sigma^{\rm D}_{\rm LH}(s')$, $\sigma^{\rm D}_{J/\psi}(s')$ 
and $\sigma^{\rm D}_{{\mathcal R}_{{\mathcal S}\rm up{\rm 3680}}}(s')$
are the cross sections of $e^+e^-$$\rightarrow\,${LH},
$J/\psi$$\rightarrow${hadrons}, and 
${\mathcal R}_{{\mathcal S}\rm up{3680}}$$\rightarrow${nOCH}, respectively.
Here ${\mathcal R}_{{\mathcal S}\rm up{3680}}$ indicates the states
with masses above 3.680 GeV/$c^{2}$. The cross sections for CPLMRs
are taken to be 
$\sigma^{\rm D}_{\rm LH}(s')$=$f\sigma^{\rm B}_{\mu^+\mu^-}(s')+\sigma^{\rm D}_{\rm LMRs}(s')$,
where $\sigma^{\rm B}_{\mu^+\mu^-}(s')$ is the Born cross section for 
continuum $e^+e^-$$\rightarrow$$\mu^+\mu^-$ production,
$f$ is a free parameter, $f\sigma^{\rm B}_{\mu^+\mu^-}(s')$ gives the cross section for 
continuum $e^+e^-$$\rightarrow${hadrons} production~\cite{PRL97_121801_2006}, and 
$\sigma^{\rm D}_{\rm LMRs}(s')$
is the cross section for the production of LMRs decaying into LH, 
which is determined using the zeroth-order cross sections~\cite{PDG2022}
multiplied by the vacuum-polarization correction factor 
$\frac{1}{|1-\Pi(s)|^2}$~\cite{Structure_Function,PRD74_054012_Y2006}
at energies below {2~\rm{GeV}}.
The cross sections for $J/\psi$ and ${\mathcal R}_{{\mathcal S}\rm up{3680}}$ 
decaying into nOCH are, respectively, taken as $\sigma^{\rm D}_{J/\psi}(s')$=$|A_{J/\psi}(s')|^{2}$ and
$\sigma^{\rm D}_{{\mathcal R}_{{\mathcal S}\rm up{3680} }}(s')$=$|A_{\psi(3686)}(s')$+$\sum^{3}_{1}A_{k}e^{i\phi_{k}}(s')|^{2}$,
where $k=1,2,3$ are for the $\mathcal R(3760)$ state, 
$\mathcal R(3780)$ state, and $\mathcal R$, respectively, 
and $\phi_k$ are relative phases.

In the above formulations, $A_{\mathcal S}$ are the generic decay amplitudes 
for these states, which are parameterized by
$A_{\mathcal S}(s') = 
{ \sqrt{ 12\pi  { \Gamma^{ee}_{\mathcal S} \Gamma^{\rm tot}_{\mathcal S} 
                       {\mathcal B(\mathcal S \rightarrow {\rm nOCH})} } } }/ 
{[ (s' - M^{2}_{\mathcal S}) + i(\Gamma^{\rm tot}_{\mathcal S}M_{\mathcal S})]}$,
in which $\mathcal S$ stands for the 
$J/\psi$, $\psi(3686)$, $\mathcal R(3760)$, and $\mathcal R(3780)$ states,
as well as resonance $\mathcal R$;
$M_{\mathcal S}$, $\Gamma^{\rm tot}_{\mathcal S}$ 
and $\Gamma^{ee}_{\mathcal S}$ are, respectively, the mass, 
total and electronic widths of the $\mathcal S$, and 
$\mathcal B(\mathcal S \rightarrow {\rm nOCH})$ is the decay branching fraction of the $\mathcal S$.
For the $\mathcal R(3780)$
state, the total width is set to be
energy dependent, as in Ref.~\cite{PRL97_121801_2006}.

In the fit, the masses, the total widths, the electronic widths, 
and the hadronic decay branching fractions of $J/\psi$ and $\psi(3686)$ resonances 
are fixed to those given in Ref.~\cite{PDG2022}.
The $\mathcal R(3780)$ mass and total width are fixed to  
$(3781.0\pm 1.3\pm 0.5)$ MeV/$c^2$ and $(19.3\pm 3.1\pm 0.1)$ MeV~\cite{bes2_prl_2structures},
respectively. The remaining parameters are left as free parameters in the fit.

The observed cross section is described by
$\sigma^{\rm O}_{\rm nOCH}(s)=\int_{0}^{1-(4m_{\pi}^2/s)}dx~{\sigma}^{\rm D}_{\rm nOCH}[s(1-x)]~{\mathcal F(x,s)}$,
where $\mathcal F(x,s)$ is the sampling function~\cite{Structure_Function,PRL97_121801_2006},
and $m_{\pi}$ is the pion mass. The cross section as a function of energy is determined 
by a fit to $\sigma^{\rm o}_{\rm nOCH}(s)$ at the 42 energy points.
Using $\sigma^{\rm D}_{\rm nOCH}(s)$ obtained from the parameter values, 
we determine the ISR correction factors
$[1+\delta(s)]=\sigma^{\rm O}_{\rm nOCH}(s)/\sigma^{\rm D}_{\rm nOCH}(s)$
at the 42 energy points, which are shown in Fig.~\ref{fig:eff_IstCrrF_nOCH} (bottom).
Dividing the observed nOCH cross section $\sigma^{\rm o}_{\rm nOCH}(s)$ 
by $(1+\delta(s))$ yields the observed dressed nOCH cross section $\sigma^{\rm d}_{\rm nOCH}(s)$, 
where the lower case superscripts are used to distinguish the measured 
from the predicted 
(upper case superscripts)
quantities. 

\begin{figure}[htbp]
 \centering
\includegraphics[width=0.49\textwidth,height=0.38\textwidth]{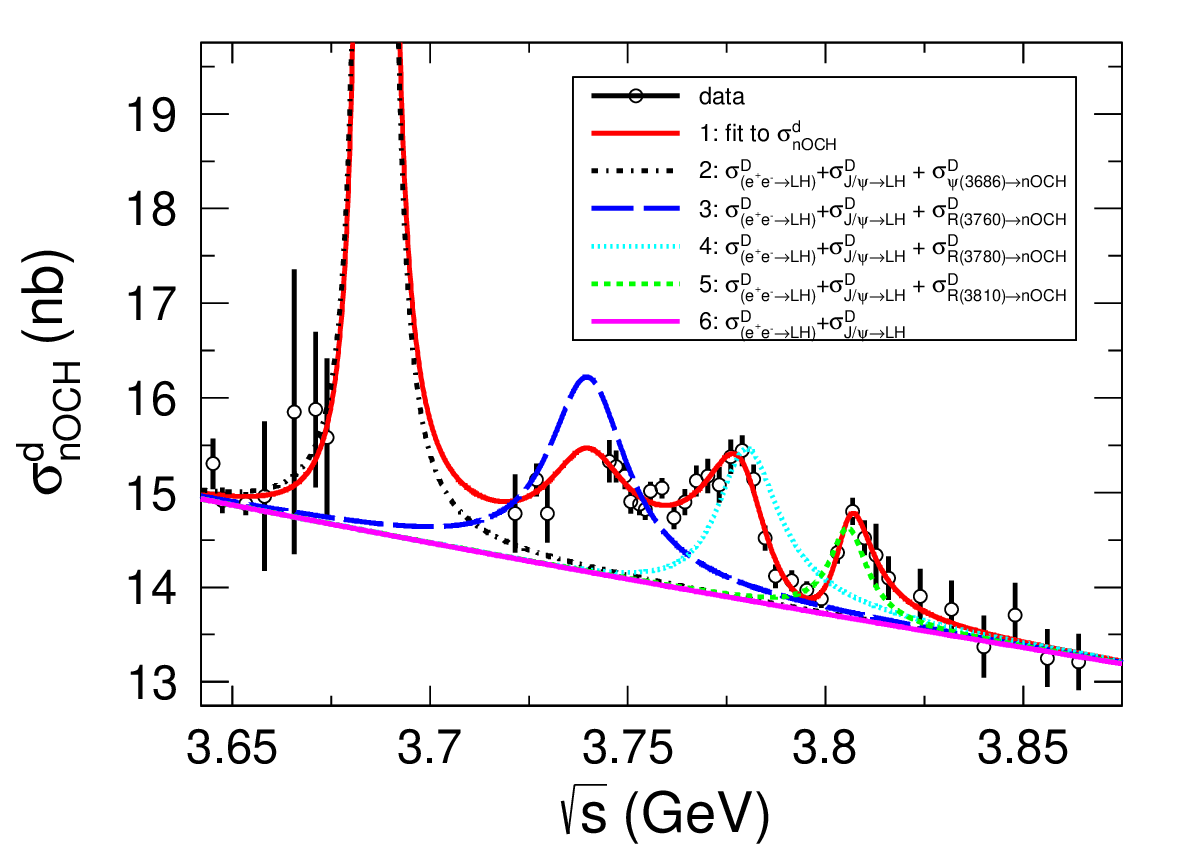}
 \caption{
The dressed cross sections for $e^+e^-\rightarrow${nOCH}, showing also 
the Result-I fit (line 1) 
and the contributions to this fit including separate contributions to 
$\sigma^{\rm D}_{{\mathcal R}_{{\mathcal S}\rm up{3680}}}(s)$ (lines 2--6). }
 \label{fig:fit_iterative_dress_CS_R3760Psi3770S3810}
\end{figure}

The circles with error bars in Fig.~\ref{fig:fit_iterative_dress_CS_R3760Psi3770S3810} 
show the $\sigma^{\rm d}_{\rm nOCH}(s)$ measurements,
where the uncertainties are  statistical.
Using the same fit procedure described above, we fit $\sigma^{\rm d}_{\rm nOCH}(s)$ 
with the function $\sigma^{\rm D}_{\rm nOCH}(s)$ given in Eq.~(\ref{eq:exp_D_xs_nOCH}),
with $x$ fixed to zero. The fit converges at two sets of acceptable parameter values.
We denote these as Result I and Result II.
Table~\ref{tab:Results_FitNOCHxs_CntPsi3686R3760Psi3770S3810}
summarizes the parameter values, where the first uncertainties 
are from the fit to the $\sigma^{\rm d}_{\rm nOCH}(s)$, and the second are systematic.
The two sets of results have the fit quality of $\chi^2/{\rm ndof}$ 
of $22.1/31$, and $22.0/31$, respectively.
As the mass of the fitted resonance is close to 3810 MeV, we denote it as $\mathcal R(3810)$.
The cross section described by Result I is superimposed on the fit results 
in Fig.~\ref{fig:fit_iterative_dress_CS_R3760Psi3770S3810}. 
Also shown are the fit results including one contribution to 
$\sigma^{\rm D}_{{\mathcal R}_{{\mathcal S}\rm up{3680}}}(s)$
included at a time. The measured mass and total width of the $\mathcal R(3760)$ state 
are consistent within $1.8\sigma$ and $0.9\sigma$, respectively, with those measured
by the BES-II experiment~\cite{bes2_prl_2structures}.
\begin{table}
\centering
\caption{Results of the fit to the cross sections for $e^+e^-\rightarrow${nOCH} 
showing the values of the mass $M_{k}$ [MeV/$c^2$],
total width $\Gamma^{\rm tot}_{k}$ [MeV], 
the product of electronic width ($\Gamma^{ee}_{k}$) 
and nOCH branching fraction (${\mathcal B}_{k}$)
$\Gamma^{ee}_{k}{\mathcal B}_{k}$ [eV],
and relative phase $\phi_{k}$ [degree],
where $k$ represents $\mathcal R(3760)$, $\mathcal R(3780)$, and $\mathcal R(3810)$.
${\mathcal B_k}$ is the branching fractions [$\%$] for $\mathcal R(3760)$ and $\mathcal R(3780)$
decays into nOCH and OC.}
\begin{tabular}{llr}
\hline\hline
Parameters                                                          &      Result I                              &  Result II             \\
\hline
$M_{\mathcal R(3760)}$
                                                                    & $3739.9 \pm  4.2 \pm 2.6$                   & $3739.7\pm 3.9 \pm 2.6$    \\
$\Gamma^{\rm tot}_{\mathcal R(3760)}$
                                                                    &   $ 23.9 \pm 8.2 \pm 4.8$                   & $22.5\pm  8.3 \pm 4.5$     \\
$\Gamma^{ee}_{\mathcal R(3760)}{{\mathcal B}_{\mathcal R(3760)}}$
                                                                    & $ 46.8 \pm 29.9 \pm 25.1$                   & $11.9 \pm   9.0 \pm 6.4$   \\
$\phi_{\mathcal R(3760)}$
                                                                    &  $ 228 \pm 52 \pm 58$                       & $113\pm   51  \pm 29$    \\
$\Gamma^{ee}_{\mathcal R(3780)}{{\mathcal B}_{\mathcal R(3780)}}$
                                                                    &   $29.9 \pm 16.1 \pm 3.7$                   & $25.3 \pm 11.6 \pm 3.1$     \\
$\phi_{\mathcal R(3780)}$
                                                                    &  $82 \pm 126 \pm 17$                        & $250\pm 119  \pm 52$       \\
${\rm M}_{\mathcal R(3810)}$
                                                                    &  $3805.7 \pm  1.1 \pm 2.7$                  & $3805.7\pm  1.1 \pm  2.7$    \\
${\rm \Gamma^{\rm tot}_{\mathcal R(3810)} }$
                                                                    &  $ 11.6 \pm 2.9 \pm  1.9$                   & $11.5\pm  2.8 \pm 1.9$   \\
$\Gamma^{ee}_{\mathcal R(3810)}{{\mathcal B}_{\mathcal R(3810)}}$
                                                                    & $10.9 \pm 3.8 \pm 2.5$                      &  $11.0 \pm 3.4 \pm 2.5$   \\
$\phi_{\mathcal R(3810)}$
                                                                    &   $ 52 \pm 149 \pm 25$                      & $215\pm 148  \pm 103$     \\
$f$                                                                 & $2.28\pm 0.01\pm    0.07$                   & $2.28\pm 0.01\pm 0.07$        \\
%                                                                   &                                             &                               \\
${\mathcal B}[\mathcal R(3760)\rightarrow$${\rm nOCH}]$             & $25.2\pm 16.1 \pm 30.4$                     & $6.4\pm 4.8 \pm 7.7$        \\
${\mathcal B}[\mathcal R(3780)\rightarrow$${\rm nOCH}]$             & $12.3\pm 6.6 \pm 8.3$                       & $10.4\pm 4.8 \pm 7.0$        \\
${\mathcal B}[\mathcal R(3760)\rightarrow$${\rm OC}]$               & $74.8\pm 16.1 \pm 30.4$                     & $93.6\pm 4.8 \pm 7.7$        \\
${\mathcal B}[\mathcal R(3780)\rightarrow$${\rm OC}]$               & $87.7\pm 6.6 \pm 8.3$                       & $89.6\pm 4.8 \pm 7.0$       \\
\hline\hline
\end{tabular}
\label{tab:Results_FitNOCHxs_CntPsi3686R3760Psi3770S3810}
\end{table}

To estimate the systematic uncertainties of the fitted parameter 
in Table~\ref{tab:Results_FitNOCHxs_CntPsi3686R3760Psi3770S3810},
we vary the values of the $\sigma^{\rm d}_{\rm nOCH}(s)$, the values of 
the $\sigma^{\rm o}_{D\bar{D}}(s)$,
and the fixed parameters by $\pm 1\sigma$, refit the $\sigma^{\rm d}_{\rm nOCH}(s)$, 
and take the difference between the refitted
parameter value and 
the baseline fit result as the corresponding systematic uncertainty. 
The estimation of the systematic uncertainty due to the uncertainty of the c.m. energies
is similar to that described in Ref.~\cite{PRL127_082002_Y2021}.
Adding these uncertainties in quadrature yields the total systematic uncertainty.

Dividing the measured values for
$\Gamma^{ee}_{\mathcal R(3760)}{\mathcal B}_{\mathcal R(3760)}$
and 
$\Gamma^{ee}_{\mathcal R(3780)}{\mathcal B}_{\mathcal R(3780)}$
in Table~\ref{tab:Results_FitNOCHxs_CntPsi3686R3760Psi3770S3810}
by 
$\Gamma^{ee}_{\mathcal R(3760)}=(186\pm 201\pm 8){~\rm eV}$~\cite{bes2_prl_2structures}
and
$\Gamma^{ee}_{\mathcal R(3780)}= (243\pm 160\pm 9)~{\rm eV}$~\cite{bes2_prl_2structures},
respectively, yields the nOCH branching fractions for the decays 
$\mathcal R(3760) \rightarrow{\rm nOCH}$ and $\mathcal R(3780) \rightarrow{\rm nOCH}$, 
which are shown in Table~\ref{tab:Results_FitNOCHxs_CntPsi3686R3760Psi3770S3810}. 
Their corresponding OC branching fractions are also presented 
in Tab.~\ref{tab:Results_FitNOCHxs_CntPsi3686R3760Psi3770S3810}.
These nOCH branching fractions of $\mathcal R(3780)$ decays 
are in good agreement  with 
$\mathcal B[\psi(3770)\rightarrow${non}-$D\bar{D}$$)]=(15.1 \pm 5.6 \pm 1.8)\%$~\cite{PLB641_145_2006,
PRL97_121801_2006,PRD76_122002_2007,PLB659_74_2008}
measured by the BES-II experiment.

By removing  the $\mathcal R(3810)$ from the
$\sigma^{\rm D}_{{\mathcal R}_{{\mathcal S}\rm up{3680}}}(s)$
as discussed above, the $\chi^2/{\rm ndof}$ 
of the fit changes from $22.1/31$ to $116.1/36$,
indicating that the significance of the $\mathcal R(3810)$ signal is  $8.8 \sigma$.
Similarly, the significance of the $\mathcal R(3760)$ signal is determined to be
$8.1\sigma$ by comparing the difference of $\chi^2/{\rm ndof}$ 
relative to the number degrees of freedom with and without including the $\mathcal R(3760)$ component 
in the fit. A similar procedure is applied to determine the $\mathcal R(3780)$
significance, which is $13.7\sigma$.
These significances include the systematic uncertainties.

The charmonium model~\cite{Eithtin_chmonuim_prd1978} 
predicts only the $1^3D_1$ state existing in the c.m. energy range from 3.733 to 3.870 GeV, 
which is generally assumed to be the $\psi(3770)$,  and so 
the $\mathcal R(3760)$ and $\mathcal R(3810)$ state
are presumably two non-$c\bar{c}$ states. 
The $\mathcal R(3760)$ state
can be explained as a $p$-wave resonance of a four-quark
($c\bar{c}q\bar{q}$) state~\cite{A_De_Rujule_PRL38_317_Year1977, Rafe_Hyam_Schindler_SLAC219}. 
It can be thought of either as an OC molecular state, or as a four-quark 
bound state~\cite{Rafe_Hyam_Schindler_SLAC219}.
Reference~\cite{S_Dubynskiy_M_B_Voloshin_PRD78_116014_Y2008} interprets
$\mathcal R(3760)$ as a possible molecular OC threshold resonance.
The most salient features of the $\mathcal R(3760)$ state
is that its mass, which is  
$(3739.9\pm 4.2\pm 2.6)$ MeV/$c^2$ for Result I [$3739.7\pm 3.9\pm 2.6)$ MeV/$c^2$ for Result II], is just at the $D^+D^-$ 
threshold $(3739.3\pm 0.1)$ MeV/$c^2$, and its branching fraction is 
${\mathcal B}[\mathcal R(3760)\rightarrow{\rm OC}]=(74.8\pm 16.1 \pm 30.4)\%$
[${\mathcal B}[ \mathcal R(3760)\rightarrow{\rm OC} ]=(93.6\pm 4.8 \pm 7.7)\%$].
These experimental facts can lead one naturally to interpret the $\mathcal R(3760)$ state
as an OC pair molecular state, 
but containing a simple four-quark state component.
As no signal for the decay $\mathcal R(3810)$$\rightarrow$$D\bar{D}$ 
is observed in the cross sections for $e^+e^-$$\rightarrow$$D\bar{D}$ 
at c.m. energies around 3.810 GeV~\cite{BaBar_PRD76_111105_Y2007, Belle_PRL77_011103_Y2008}, 
and the $\mathcal R(3810)$ mass $(3805.7\pm 1.1\pm 2.7)$ MeV/$c^2$ for Result I 
[$(3805.7\pm 1.1\pm 2.7)$ MeV/$c^2$ for Result II] is exactly at 
the ${h_c}\pi^+\pi^-$ threshold $(3804.5\pm 0.1)$ MeV/$c^2$, 
the $\mathcal R(3810)$ state can be interpreted 
as a hadrocharmonium resonance~\cite{Voloshin_PLB66_344_y2008}.

In summary, we have measured the inclusive nOCH cross sections of $e^+e^-$$\rightarrow${~nOCH}
with improved precision at c.m.\ energies from 3.645 to 3.871 GeV. 
We observe three resonances: $\mathcal R(3760)$, $\mathcal R(3780)$, and 
$\mathcal R(3810)$ in the energy-dependent line shape of the nOCH cross sections with significances 
of $8.1\sigma$, $13.7\sigma$, and $8.8\sigma$,
respectively. The $\mathcal R(3760)$ and $\mathcal R(3780)$ states
are observed for the first time in the nOCH cross sections, while the $\mathcal R(3810)$ 
state is observed for the first time
with  mass $(3805.7 \pm 1.1 \pm 2.7)$~[$(3805.7 \pm 1.1 \pm 2.7)$]~MeV/$c^2$,
total width $(11.6 \pm 2.9 \pm 1.9)$~[$(11.5 \pm 2.8 \pm 1.9)$]~MeV, and
the product of the electronic width and the nOCH decay branching fraction 
$(10.9\pm 3.8\pm 2.5)$~[$(11.0\pm 3.4\pm 2.5)$]~eV for Result I [Result II].
In addition, for the first time, we measure 
${\mathcal B}[\mathcal R(3760)$$\rightarrow${\rm nOCH}]=$(25.2\pm 16.1 \pm30.4)\%[(6.4\pm 4.8 \pm 7.7)\%]$
and 
${\mathcal B}[\mathcal R(3780)$$\rightarrow${\rm nOCH}]=$(12.3\pm 6.6 \pm 8.3)\%[(10.4\pm 4.8 \pm 7.0)\%]$.
The $\mathcal R(3760)$ state
can be interpreted as an OC pair molecular state,
but containing a simple four-quark state component.
The $\mathcal R(3810)$ state can be interpreted 
as a hadrocharmonium state~\cite{Voloshin_PLB66_344_y2008}.

\vspace{-0mm}
The BESIII Collaboration thanks the staff of BEPCII and the IHEP computing center for their strong support. 
This work is supported in part by National Key R$\&$D 
Program of China under Contracts 
No. 2009CB825204, 
No. 2020YFA0406300, 
No. 2020YFA0406400; 
National Natural Science Foundation of China (NSFC) under Contracts 
No. 10935007, 
No. 11635010, 
No. 11735014, 
No. 11835012, 
No. 11935015, 
No. 11935016, 
No. 11935018, 
No. 11961141012, 
No. 12022510, 
No. 12025502, 
No. 12035009, 
No. 12035013, 
No. 12061131003, 
No. 12192260, 
No. 12192261, 
No. 12192262, 
No. 12192263, 
No. 12192264, 
No. 12192265, 
No. 12221005, 
No. 12225509, 
No. 12235017; 
the Chinese Academy of Sciences (CAS) Large-Scale Scientific Facility Program; 
the CAS Center for Excellence in Particle Physics (CCEPP); 
CAS Key Research Program of Frontier Sciences under Contracts 
No. QYZDJ-SSW-SLH003, 
No. QYZDJ-SSW-SLH040; 
100 Talents Program of CAS; 
The CAS Research Program under Code No. Y41G1010Y1;
The CAS Other Research Program under Code No. Y129360;
The Institute of Nuclear and Particle Physics (INPAC) and Shanghai Key Laboratory for Particle Physics and Cosmology; 
ERC under Contract No. 758462; European Union's Horizon 2020 research and innovation programme 
under the Marie Sklodowska-Curie grant agreement under Contract No. 894790; 
German Research Foundation DFG under Contracts 
No. 443159800, 
No. 455635585, 
Collaborative Research Center CRC 1044, FOR5327, GRK 2149; 
Istituto Nazionale di Fisica Nucleare, Italy; Ministry of Development of Turkey under Contract No. DPT2006K-120470; 
National Research Foundation of Korea under Contract No. NRF-2022R1A2C1092335; National Science and Technology fund of Mongolia; 
National Science Research and Innovation Fund (NSRF) via the Program Management Unit 
for Human Resources 
$\&$
Institutional Development, Research and Innovation of Thailand under Contract No. B16F640076; 
Polish National Science Centre under Contract No. 2019/35/O/ST2/02907; The Swedish Research Council; 
U. S. Department of Energy under Contract No. DE-FG02-05ER41374.

\end{document}